\documentstyle[11pt,newpasp,twoside]{article}
\def\edcomment#1{\iffalse\marginpar{\raggedright\sl#1\/}\else\relax\fi}
\reversemarginpar

\begin{document}
\title{Galactic Evolution of Beryllium and Oxygen}
 \author{Garik Israelian$^{1,2}$, Ram\'on J. Garc\'\i a L\'opez$^{1,2}$ and 
 Rafael Rebolo$^{1,3}$}
\affil{1.- Instituto de Astrof\'\i sica de Canarias, E-38200 La Laguna, \\
           \ Tenerife, Spain\\
       2.- Departamento de Astrof\'\i sica, Universidad de La Laguna,
           Av. \\ 
	   \ Astrof\'\i sico Francisco S\'anchez s/n, E-38071 La Laguna, 
	   Tenerife, Spain \\
       3.- Consejo Superior de Investigaciones Cient\'\i ficas, Spain}

\begin{abstract}

We discuss the early evolution of beryllium and oxygen in our Galaxy by
comparing abundances of these elements for halo and disk metal-poor stars.
Both, O and Be rise as we go progressively to more metal-rich stars, showing a
slope $0.41\pm 0.09$ ([Be/O] vs [Fe/H]) for stars with [Fe/H]$\leq -1$. This
relationship provides an observational constraint to the actually proposed 
Galactic Cosmic Ray theories.

\end{abstract}

\section{Introduction}

First attempts to measure beryllium abundances in metal-poor stars by  Molaro
\& Beckman (1984) and  Molaro, Beckman \& Castelli (1984) demonstrated that
stars in the early Galaxy formed with much lower Be abundances than in the
present epoch. First detection of Be in metal-poor stars was achieved by Rebolo
et al. (1988) and further studies by S. Ryan, G. Gilmore, A. Boesgaard, P.
Molaro, R. J. Garc\'\i a L\'opez and their respective collaborators revealed a
clear linear correlation  with iron. 

Accelerated protons and $\alpha$-particles in cosmic rays interact with ambient
CNO in ISM and create Be. According to the standard Galactic Cosmic Ray (GCR)
theory, these interactions in the general ISM should have given a quadratic
relation between Be and O. Alternatively, spallation of cosmic ray CNO nuclei
accelerated out of freshly  processed material could account for the primary
character of the observed early galactic evolution of Be. Another production
site is  the collective acceleration by SN shocks of ejecta-enriched matter in 
the interiors of superbubbles. In these two cases, the evolution of Be should 
reflect the production of CNO from massive stars.  Oxygen is mostly produced by
Type II SNe while iron is produced in both, Type II and in Type Ia SNe. The
fact that  Type Ia SNe have longer lifetime progenitors has been commonly
used   to argue that oxygen must be overabundant in very old stars.
Observational  evidence  for high [O/Fe] ratios in many metal-poor stars has
been reported over  the last two decades.

Based on the study of [O\,{\sc i}] lines at 6300 and 6363 \AA\ in evolved stars
(though the second line at 6363 \AA\ is not visible in very metal-poor stars
and the analysis is based {\it only on one line}), several authors have found 
that [O/Fe]$=0.3-0.4$  dex at [Fe/H]$< -1$ and is constant until [Fe/H]$\sim
-3$ (e.g. Barbuy 1988 and Kraft et al. 1992). In contrast with this result,
oxygen abundances derived in unevolved stars using the O\,{\sc i} IR triplet at
7774 \AA\ (Abia \& Rebolo 1989; Tomkin et al. 1992; King \& Boesgaard 1995;
and Cavallo, Pilachowski, \& Rebolo 1997) point towards linearly increasing
[O/Fe] values with decreasing  [Fe/H] and reaching a ratio $\sim 1$ for stars
with [Fe/H]$\sim -3$. This may suggest a higher production of oxygen during the
early Galaxy.

We discuss in this paper the comparison of these abundances with those derived
from OH lines located in the near-UV part of the spectra of metal-poor stars,
and their relation with beryllium abundances consistently derived from the
same spectra.

\section{Observations and Analyses}

The observations were carried out in different runs using the UES ($R=\lambda
/\Delta\lambda\sim 50000$) of the 4.2-m WHT at the Observatorio del Roque de
los Muchachos (La Palma), and the UCLES ($R\sim 60000$) of the  3.9-m AAT. The
spectral region observed spanned typically from 3080 to 3300 \AA, where
several OH lines and the Be\,{\sc ii} doublet at 3131 \AA\ are located. 
Details of the abundance analyses can be found in Garc\'\i a L\'opez,
Severino, \& Gomez (1995) and Israelian, Garc\'\i a L\'opez, \& Rebolo (1998)
for beryllium and oxygen, respectively. The stellar parameters play an
important role in the abundance determinations from near-UV lines. Effective
temperatures ($T_{\rm eff}$) for our stars were estimated using the Alonso et
al. (1996) calibrations versus $V-K$ and $b-y$ colors, which were derived by
applying the infrared flux method (IRFM), and cover a wide range of spectral
types and metal content. These temperatures were used to compute synthetic
spectra around the Be\,{\sc ii} doublet and slightly modified, within the
error bars provided by the calibrations, until obtaining a good reproduction
of this region in the observed spectra. Metallicities were adopted from
literature values obtained from high resolution spectra. Adopted gravities,
derived using the  accurate parallaxes  measured by {\it Hipparcos} (ESA
1997), are larger by 0.28 dex in average than the values adopted by Israelian
et al. (1998). This implies a mean small reduction of 0.09 dex in the oxygen
abundances inferred from the OH lines with respect to that work, which does not
affect significantly their original results.

Beryllium abundances reported here were obtained from the Be\,{\sc ii}
resonance doublet located at 3130.421 and 3131.065 \AA. The first line, which
is also the strongest one, is severely blended with atomic and molecular lines
of other species, and the abundance determination usually relies only on the
other line, more isolated and weaker. The abundances used in this work are
those presented in Garc\'\i a L\'opez (1999).

\section{Oxygen}

Israelian et al. (1998) presented  new oxygen abundances derived from near-UV
OH lines (which form in the same layers of the atmosphere as [O\,{\sc i}]) for
24 metal-poor stars.  They have concluded that the [O/Fe] ratio of metal-poor
stars increases from 0.6 to 1 between [Fe/H]=$-$1.5 and $-3$, with a slope of
$-0.31\pm 0.11$. Contrary to the previously accepted picture (see e.g. Bessell,
Sutherland, \& Ruan 1991, who used older model atmospheres with a coarser
treatment of the opacities in the UV), these new oxygen abundances derived from
low-excitation OH lines, agreed well with those derived from high-excitation
lines of the O\,{\sc i} IR triplet at 7774 \AA. The comparison with oxygen
abundances derived using O\,{\sc i} data from Tomkin et al. (1992) showed a
mean difference of $0.00\pm 0.11$ dex for the stars in common. Boesgaard et al.
(1999a) made a similar analysis of  several metal-poor stars using a different
set of OH lines. They found a very good agreement with the results obtained by
Israelian et al. (1998), and basically the same dependence of [O/Fe] versus
metallicity. This is clearly seen in the upper panel of Figure 1.

\begin{figure}
\plotfiddle{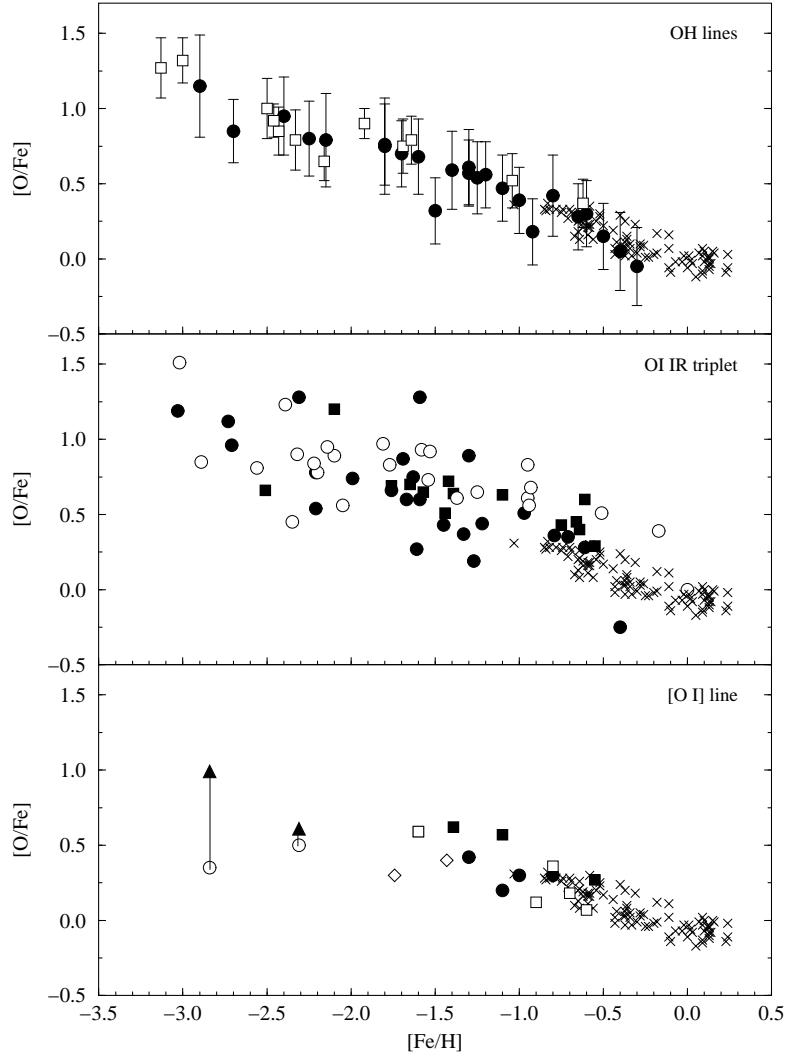}{13.5cm}{360}{55}{55}{-160}{-30}
\caption[]{[O/Fe] vs. [Fe/H] for unevolved stars. Abundances from OH lines
	were derived by Israelian et al. (1998; filled circles) and
	Boesgaard et al. (1999a; open squares, corrected to the scale of
	stellar parameters adopted by Israelian et al.). Abundances from the 
	IR triplet were derived in NLTE by Mishenina et al. (2000; filled 
	circles), Cavallo et al. (1997; filled squares, corrected for NLTE
	effects by Mishenina et al.), and in LTE by Boesgaard et al. (1999;
	open circles). Finally, abundances from the [O\,{\sc i}] line come
	from Spiesman \& Wallerstein (1991; open diamonds), Spite \& Spite 
	(1991; open squares), Israelian et al. (1998; filled circles), 
	Mishenina et al. (2000; filled squares), and Fulbright \& Kraft (1999; 
	open circles). Filled triangles indicate the change in abundances
	associated with the change in gravity according to the {\it Hipparcos}
	parallaxes for the two stars studied by Fulbright \& Kraft. The
	abundances derived by Edvardsson et al. (1993; crosses) are shown in 
	the three plots to indicate the trend in metal-rich stars.}
\end{figure}

The UV ``missing opacity'' problem discussed by Balachandran \& Bell (1998),
which could affect both oxygen and beryllium abundance determinations from these
lines, has been studied recently by Allende Prieto \& Lambert (2000). These 
authors have found a good agreement between $T_{\rm eff}$s obtained from the
IRFM and from the near-UV continuum for stars with $4000\leq T_{\rm eff}\leq
6000$ K when accurate {\it Hipparcos} gravities are used. This also agrees with
our good reproduction of the near-UV spectral region using the IRFM
temperatures. This result indicates that the model atmospheres used provide an
adequate description of the near-UV continuum forming region. In any case, even
if a not well understood opacity problem would exist as described by
Balachandran \& Bell, it would have a minor effect on the OH results since most
of the stars in the samples of Israelian et al. and Boesgaard et al. are hotter
than the Sun and very metal-poor. The corrections to oxygen abundances for
individual stars would be lower than 0.15 dex, not changing significantly the
[O/Fe] vs. [Fe/H] trend. 

A new non-LTE analysis of the O\,{\sc i} IR triplet for a sample of 38
metal-poor stars performed by Mishenina et al. (2000) gives consistent results
with those of Abia \& Rebolo (1989), Tomkin et al. (1992) and Kiselman (1993),
and indicates that the mean value of the non-LTE correction in unevolved
metal-poor stars is typically 0.1-0.2 dex. These authors confirmed the [O/Fe]
vs [Fe/H] trend discussed  by Israelian et al. (1998) and Boesgaard et al.
(1999a) from the OH lines, without finding any trend of oxygen abundances with
$T_{\rm eff}$ or $\log g$. It is also worthwhile to mention that the O\,{\sc i}
IR triplet is not affected by 3D effects, convection and small-scale
inhomogeneities in the stellar atmosphere (Asplund et al. 1999). In addition,
oxygen abundances derived form this triplet are not significantly affected by
chromospheric activity either. The central panel of Fig. 1 shows a compilation
of oxygen abundances derived using the IR triplet. The larger scatter observed
as compared with the measurements based on OH lines can be associated with the
different scales of stellar parameters ($T_{\rm eff}$, gravities, and
metallicities) adopted by the authors of each set of stars, and to the fact
that some  measurements have not been corrected for non-LTE effects. Very
recently, Carretta, Gratton, \& Sneden (2000) performed an independent analysis
of 32 metal-poor stars hotter than 4600 K using the IR triplet, and provide LTE
and non-LTE oxygen abundances which are significantly lower than those found in
previous works. A preliminary attempt to understand the reasons for this
discrepancy can be done by looking in detail into their most metal-poor star
(BD +3\deg  740, [Fe/H]$=-2.66$) where a surprisingly low oxygen abundance 
[O/Fe]$=0.38$ is claimed. A recent study of stellar parameters based on the
non-LTE analysis of iron lines (Th\'evenin \& Idiart 1999), gives a lower
effective temperature (by 140 K) and a higher gravity (by 0.3 dex) than the
values adopted by Carretta et al. for this star.  Using these latter parameters
we obtain an LTE oxygen abundance 0.4 dex higher than Carretta et al. (i.e.
[O/Fe]$_{\rm LTE}=1.05$), and for a star with these parameters the non-LTE
correction to the oxygen abundance is of the order of 0.05 dex (Mishenina et
al. 2000), much lower than the 0.25 dex value used by Carretta et al. We
therefore arrive at a value   [O/Fe]$_{\rm NLTE}\sim 1.0$, in good agreement
with the OH determination by Boesgaard et al. (1999a). Corrections to the
stellar parameters as inferred from the non-LTE analysis of Fe lines clearly
have an impact on the oxygen abundances which we will address in a forthcoming
paper.

Israelian et al. (1998) found four dwarfs in their sample for which oxygen
abundances derived using [O\,{\sc i}] were in good agreement with those derived
from OH when {\it Hipparcos} gravities are used. Several oxygen measurements
for unevolved stars based on the [O\,{\sc i}]  6300 \AA\ line are compiled in
the lower panel of Fig. 1. This figure shows a similar trend than that observed
for the abundances recently derived from forbidden lines by Carretta et al.
(2000). The presence of a linear trend of [O/Fe] versus metallicity in Fig. 1
strongly depends on the only two measurements available at [Fe/H]$\le -2$.
These two measurements have been reported by  Fulbright \& Kraft (1999) for the
subgiants BD +37\deg 1458 and BD +23\deg 3130, which were also considered by
Israelian et al. (1998) and Boesgaard et al. (1999a; only BD +37\deg 1458 in
this case). The analysis carried out by Fulbright \& Kraft  is based on
gravities derived from LTE iron ionization balance of these subgiants where it
is well known that non-LTE effects are strong (Th\'evenin \& Idiart 1999; see
also Idiart \& Th\'evenin, this conference).  Allende Prieto et al. (1999) have
shown that gravities derived using this technique in metal-poor stars do not
agree with the gravities inferred from accurate {\it Hipparcos} parallaxes. 
They find that  gravities are systematically underestimated when derived from
ionization balances and that upward corrections of $\sim 0.5$ dex can be
required at metallicities similar to those of our stars, in good agreement with
Th\'evenin \& Idiart. We remark here that any  underestimation of gravities
will also strongly underestimate the abundances inferred from the forbidden
line. For the two stars under discussion our {\it Hipparcos} based gravities
are 0.45 and 1.05 dex (for BD +37\deg 1458 and  BD +23\deg 3130, respectively)
higher than derived by Fulbright \& Kraft, and would imply corrections in the
oxygen abundances similar to those indicated in Fig. 1 (a detailed analysis
would imply also the correction for the assumed metallicities). Our conclusion
is  that the uncertainties in the gravities of these subgiants allow the
abundances inferred from the forbidden line to be consistent with those
estimated from the OH lines or the triplet. Actually, consistency with the
other oxygen indicators is achieved for the high gravities inferred from {\it 
Hipparcos} when consistent analyses are made, and this could be taken as an
indication that the high gravities are indeed the correct ones.

\section{Beryllium and oxygen}

The dependence of $\log$(Be/H) on [Fe/H] and on [O/H] (using the abundances
derived from the OH UV lines) is essentially linear (Garc\'\i a L\'opez 1999;
Boesgaard et al. 1999b), but with different slopes: $\sim 1.1$ and $\sim 1.5$,
respectively. No evidence of a primordial plateau of Be down to
$\log$(Be/H)=$-$13.5 is found. Figure 2 shows the increase of the [Be/O] ratio
with increasing metallicity and a slope of $\sim 0.4$. This relation provides
an observational constraint to the Galactic Cosmic Ray theories. Three types
of GCR models exist at present which try to explain their observed evolution.
These are 1) a pure primary GCR from superbubbles (Ramaty, this conference),
2) a hybrid model based on  GCR and superbubble accelerated particles
(Cass\'e, this conference), which could be accomplished by a pure superbubble
model (Parizot \& Drury, this conference), and 3) standard GCR (Olive, this
conference). Apparently all these models can be adopted for both, variable and
flat [O/Fe]. However, models presented by R. Ramaty and K. Olive show more
consistency when variable [O/Fe] is adopted. 

\begin{figure}
\plotfiddle{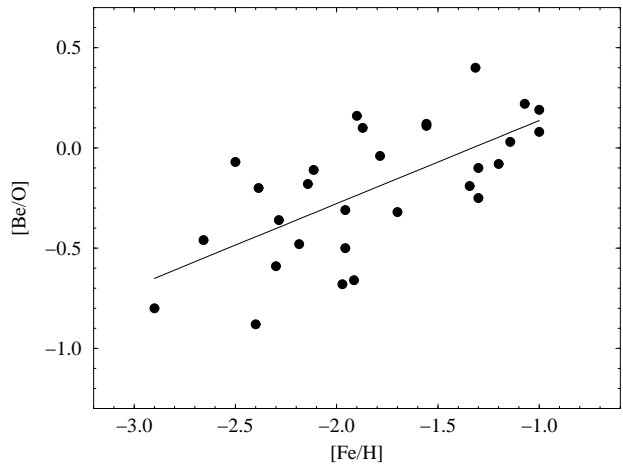}{5.2cm}{270}{35}{35}{-140}{+190}
\caption[]{[Be/O] vs. [Fe/H] for unevolved metal-poor stars. Oxygen abundances
used were derived from OH UV lines. The slope observed in the figure, $0.41\pm
0.09$, provides an observational constraint to the GCR models.}
\end{figure}

Chemical evolution models of the early Galaxy where stellar lifetimes  are
taken into account and assuming that Type Ia SN appear at a Galactic age of 30
million years can also explain the evolution of oxygen delineated in Fig. 1.
(Chiappini et al. 1999.). The evolution of  oxygen proposed in this paper also
helps to understand the evolution  of $^6$Li versus [Fe/H] and the $^6$Li/Be
ratio at low metallicities  in the framework of standard Galactic Cosmic Ray
Nucleosynthesis (Fields \& Olive 1999). In addition, Ramaty et al. (1999) have
proposed that a delay between the effective deposition times into the ISM of 
Fe and O (only a fraction of which condensed in oxide grains) can  explain a
linear trend of [O/Fe].

It has been suggested (Vangioni-Flam \& Cass\'e, this conference) to  use
magnesium as  metallicity indicator instead of oxygen. However, given the
existence of unevolved halo stars with negative [Mg/Fe] ratios (McWilliam 1997;
Carney et al. 1997), this approach may not lead to better results.  For
example, the  subdwarf BD+3\deg 740 has [O/Fe]$\sim 1$ (see previous Section)
while its [Mg/Fe]$=-0.28$ (Fuhrmann et al. 1995). Yield of Mg depends on the
extent of mixing (Argast et al. 2000) and its primordial abundance can be
changed due to the operation of the MgAl cycle.

\end{document}